\documentclass[12pt,a4paper]{article}
\usepackage{graphicx}
\usepackage{times}
\newcommand{\exosat}{{\it EXOSAT}}

\newcommand{\chandra}{{\it CHANDRA}}
\newcommand{\rxte}{{\it RXTE}}
\newcommand{\asca}{{\it ASCA}}
\newcommand{\rosat}{{\it ROSAT}}

\newcommand{\xmm}{{\it XMM}}

\newcommand{\ec}{$\eta$~Car}

\newcommand{\kms}{km~s$^{-1}$}

\newcommand{\mnras}{MNRAS}

\title{\bf The X-ray Lightcurve of WR~140}
\author{M. F. Corcoran$^{1,2}$, A. M. T. Pollock$^{2}$, K. Hamaguchi$^{1,4}$, C. Russell$^{5}$\\
\vspace{1cm}\\
\normalsize $^1$ CRESST/NASA-GSFC, Greenbelt, MD, USA\\ 
\normalsize $^2$ Universities Space Research Association, Columbia, MD, USA \\
\normalsize $^3$ European Space Agency, Villanueva de la Ca\~{n}ada, Madrid, Spain\\
\normalsize $^4$ University of Maryland, Baltimore County, Baltimore, MD, USA\\
\normalsize $^5$ University of Delaware, Baltimore, MD, USA\\
\\
\\
\normalsize Published in proceedings of \\
\normalsize"Stellar Winds in Interaction", editors T. Eversberg and J.H. Knapen. \\ 
\normalsize Full proceedings volume is available on http://www.stsci.de/pdf/arrabida.pdf
}
\date{\mbox{}}
\begin{document}
\maketitle
%
%
\def\bull{\vrule height .9ex width .8ex depth -.1ex}
\makeatletter
\def\ps@plain{\let\@mkboth\gobbletwo
\def\@oddhead{}\def\@oddfoot{\hfil\tiny\bull\quad
Workshop ``Stellar Winds in Interaction'' Convento da Arr\'abida, 2010 May 29 - June 2\quad\bull}%
\def\@evenhead{}\let\@evenfoot\@oddfoot}
\makeatother

%
%
\def\beginrefer{\section*{References}%
\begin{quotation}\mbox{}\par}
\def\refer#1\par{{\setlength{\parindent}{-\leftmargin}\indent#1\par}}
\def\endrefer{\end{quotation}}
%
%
{\noindent\small{\bf Abstract:} 
WR~140 is a canonical massive ``colliding wind'' binary system in which periodically-varying X-ray emission is produced by the collision between the wind of the WC7 and O4-5 star components in the space between the two stars.  We have obtained X-ray observations using the \rxte\ satellite observatory through almost one complete orbital cycle including two consecutive periastron passages.  We discuss the results of this observing campaign, and the implications of the X-ray data for our understanding of the orbital dynamics and the stellar mass loss.
}
%
%
\section{Introduction: WR~140 as a ``Canonical'' System}
Although massive stars are the most important objects in the Universe for generation of metals needed for the formation of rocky planets and biological entities, there remain many open questions about these objects, especially concerning how they evolve and die.  Understanding this evolutionary process is largely a matter of understanding how these massive stars lose mass and angular momentum.  The study of massive binaries is one key since in these systems the fundamental stellar parameters can be directly measured (at least for those fortuitously situated) and because wind-wind (or even wind-star) collisions in these systems provide an in-situ measure of the radiatively driven process which is the major means of mass and angular momentum loss for most of a massive star's natural life.  

\begin{figure}[htbp] 
   \centering
    \includegraphics[width=3.5in]{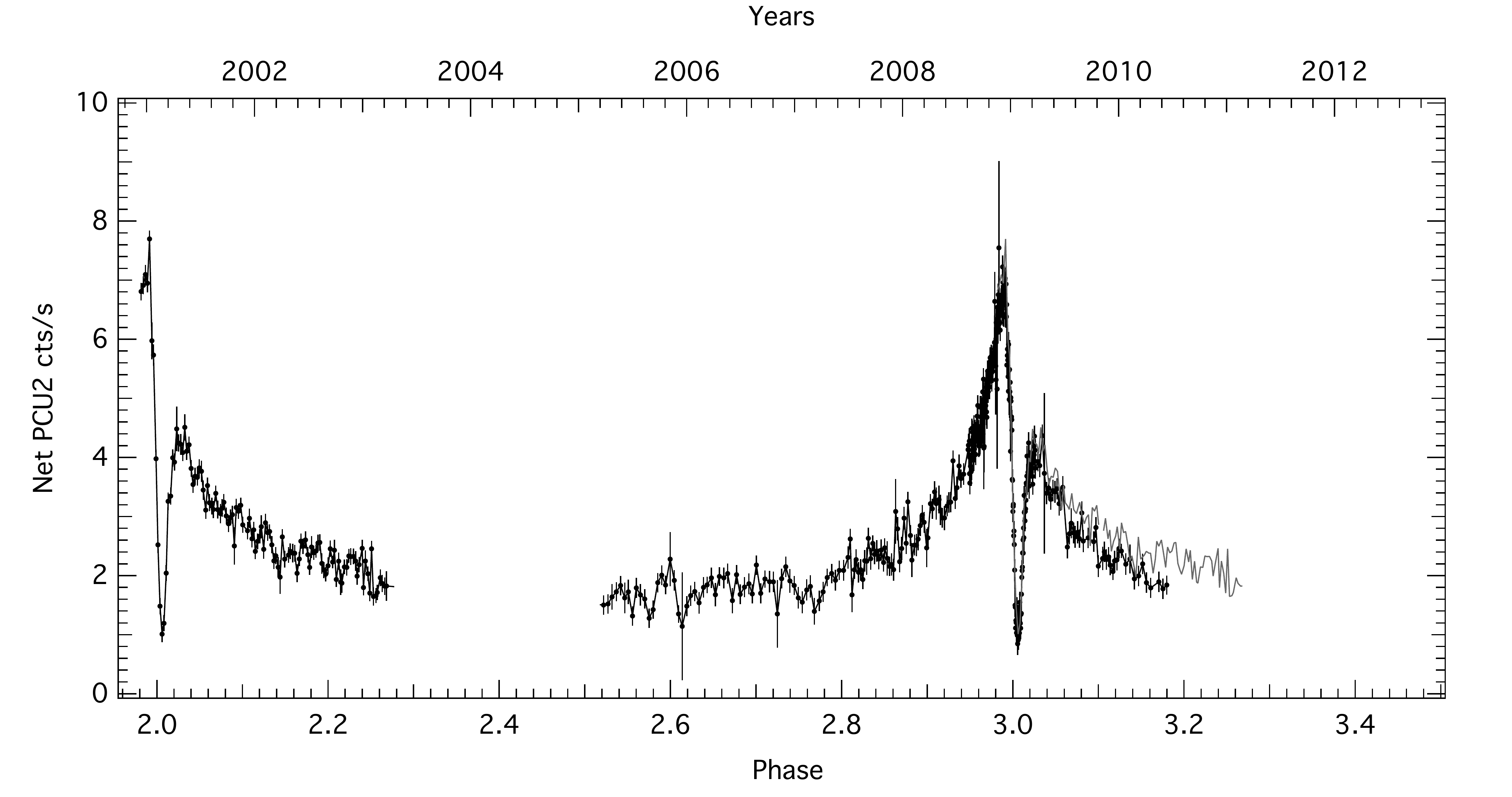}
 \includegraphics[width=2.8in]{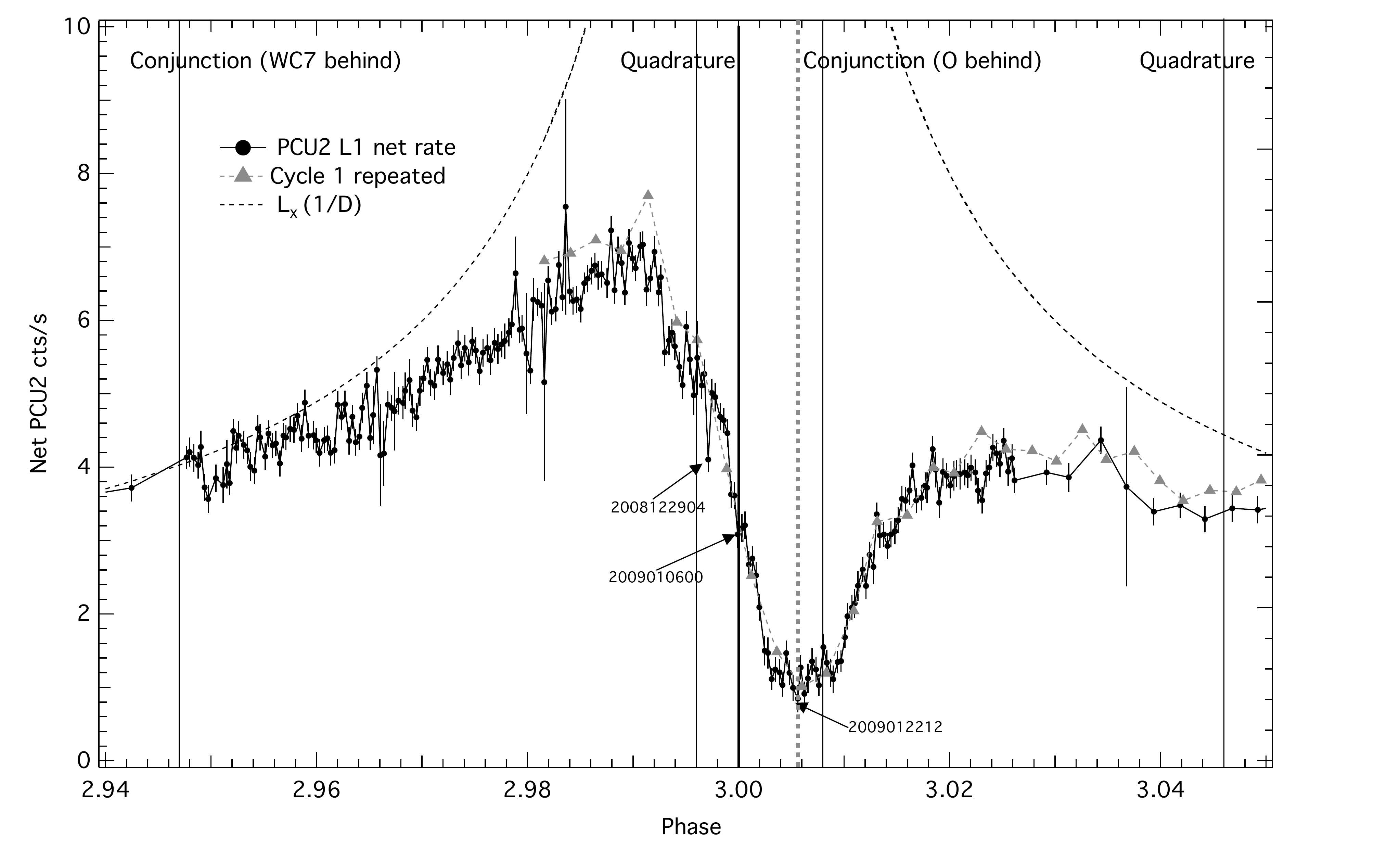}
  \caption{\footnotesize{\textit{Left}: The \rxte\ lightcurve of WR~140, December 2000 to June 2010.  \textit{Right}: Detailed view of the two minimum observed by \rxte. The data in grey represents is the \rxte\ data offset by one orbital period, adopted to be $P=2897$ days. }}
   \label{fig:pcu2}
\end{figure}

WR~140 is often termed a ``canonical'' colliding wind system by stellar astrophysicists.  Presumably this is meant in the sense that WR~140 may be used to establish the rule of behaviour of such massive binaries. Physicists refer to canonical pairs of complementary variables and this is at least superficially appropriate for WR~140. A canonical object also suggests a kind of systemic uniqueness, and indeed WR~140 as a system is nearly unique (if uniqueness can be approached in degrees): it's an evolved binary system with an unusually long (7.9 year) period, and it possesses nearly the highest eccentricity of all known stellar orbits (see Williams et al. 1990, and R.~Fahed et al., these proceedings). It is well beyond the scope of the current discussion to try to understand how this binary collapsed into this peculiar state (stellar capture?  explosive disintegration of a third component?).  Nature, however, has kindly provided this laboratory in which the stellar separations and orbital velocities vary by an order of magnitude around the orbit.  This allows the patient astronomer to measure the dependence of the state of the shock heated gas in the wind-wind collision on the local thermodynamic variables, and to look for the influence of subtle (and sometimes not so subtle) effects due to radiative transfer of photospheric photons winding their way through a complex interacting wind structure, or due to the sudden onset of fluid instabilities.  All these things make WR~140 an object of intense interest for observational astronomers of all stripes, stellar evolutionists, hydrodynamicists, and plasma physicists.
 
\section{X-ray Emission from Colliding Wind Binaries Like WR~140}

The X-ray band is extremely useful for the study of colliding wind shocks in general and for the study of WR~140 in particular.  In WR~140 the stellar winds have terminal speeds $V_{\infty}\sim 3000$~\kms.  A collision at these speeds  produces shock-heated gas at a temperature of 
$$T_{\rm s} \approx\frac{3}{16} \mu V_{\infty}^{2}R^{-1}\approx 1.51 \times10^{5} (V_{\infty}/100~\mbox{\kms})^{2}\,\mbox{K} \approx 1.3\times10^{8}\,\mbox{K}$$ 
(where $R$ is the gas constant and $\mu$ the mean molecular weight of the winds), suitably emitting in the X-ray band at energies of 1--10 kilo-electron volts (keV). Because a substantial portion of the stellar winds becomes involved in the shock, the amount of X-ray emission generated is substantial and observable with modern X-ray satellite observatories.  WR~140 was first detected as an X-ray bright source on 1984 May 19  by the \exosat\ observatory (Pollock 1985), and has been observed many times subsequently by X-ray observatories like \rosat, \asca, \xmm, \chandra, and \rxte.

\section{\rxte\ Observations}

The \textit{Rossi X-ray Timing Explorer} (\rxte, Bradt et al. 1993), launched in December 1995, is a satellite observatory that flexibly observes variable X-ray emitting cosmic objects on timescales of microseconds to years. The workhorse instrument, the Proportional Counter Array (PCA), is sensitive to X-rays emitted in the 2--60 keV band, which includes the 10--20 million K emission generated within WR~140's wind-wind collision shock. The first observation of WR~140 by \rxte\ was obtained on 2000 December 9, just 53 days before periastron passage, and \rxte\ observed the system about once per week up to about 778 days past periastron passage.  Observations resumed on 2005 March 8, just past apastron of the stellar orbits, and continue up through the time of this writing, at a variable cadence of a few observations per month to one observation per day during the most recent periastron passage in 2009 January.
\begin{figure}[htbp] 
   \centering
   \includegraphics[width=5in]{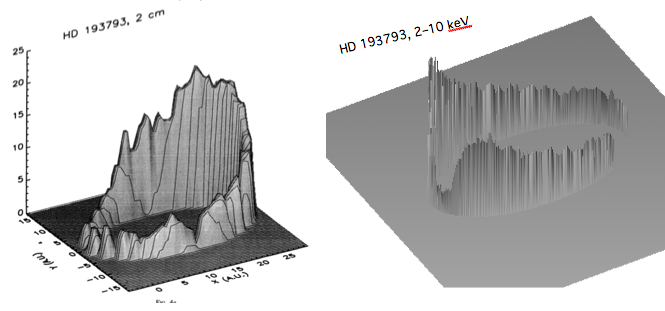} 
   \caption{\footnotesize{Comparison of the orbital variation of the 2\,cm radio flux (from White \& Becker 1995) and the 2--10 keV X-ray flux.}}
   \label{fig:xsurf}
\end{figure}


Figure \ref{fig:pcu2} shows the WR~140 \rxte\ lightcurve as a function of phase $\phi$ (where $\phi=0$ is periastron passage) and time in years.  The general characteristic is a gradual increase in X-ray brightness as the stars approach periastron, followed by a rapid decline to a brief minimum state, a quick recovery followed by  a gradual decline in brightness. It's worth noting that there is no rapid ($\sim$weeks) variations (``flaring'') of the X-ray brightness as seen in WR~140's ``sister star'' \ec\ just prior to the start of its X-ray minimum. As shown in Fig.~\ref{fig:pcu2}, the X-ray minima from the two orbits observed by \rxte\ agree very well with each other in terms of brightness variability, depth and duration.  There may, however, be slight variations in the level of X-ray brightness between the two cycles.  Figure \ref{fig:pcu2} also shows a curve which represents a $1/D$ variation, where $D$ is the separation between the two stars.  Near X-ray minimum the X-ray brightness curve does not follow a $1/D$ variation as it would if the colliding wind shock were adiabatic and there were no line-of-sight absorption variations.  We also note that X-ray minimum apparently precedes O-star conjunction (i.e. the time in the orbit when the O star is behind the WR star) by about 6 days.  Figure \ref{fig:xsurf} compares the orbital variation of the 2\,cm radio flux with the 2--10\,keV X-ray flux.  As can be seen from this figure, both the radio flux and X-ray flux show minima near periastron passage/O star conjunction when the colliding wind shock is viewed through the thick wind of the WC7 star. The radio decline is much more gradual than the X-ray minimum and in fact just before periastron passage the radio emission is in its minimum state while the X-ray emission is still increasing. The X-ray recovery from minimum is much more rapid than the radio recovery.   The radio emission reaches a maximum intensity near apastron when the colliding wind shock moves out from behind the radio photosphere of the system.

\begin{figure}[htbp] 
   \centering
   \includegraphics[width=3.5in]{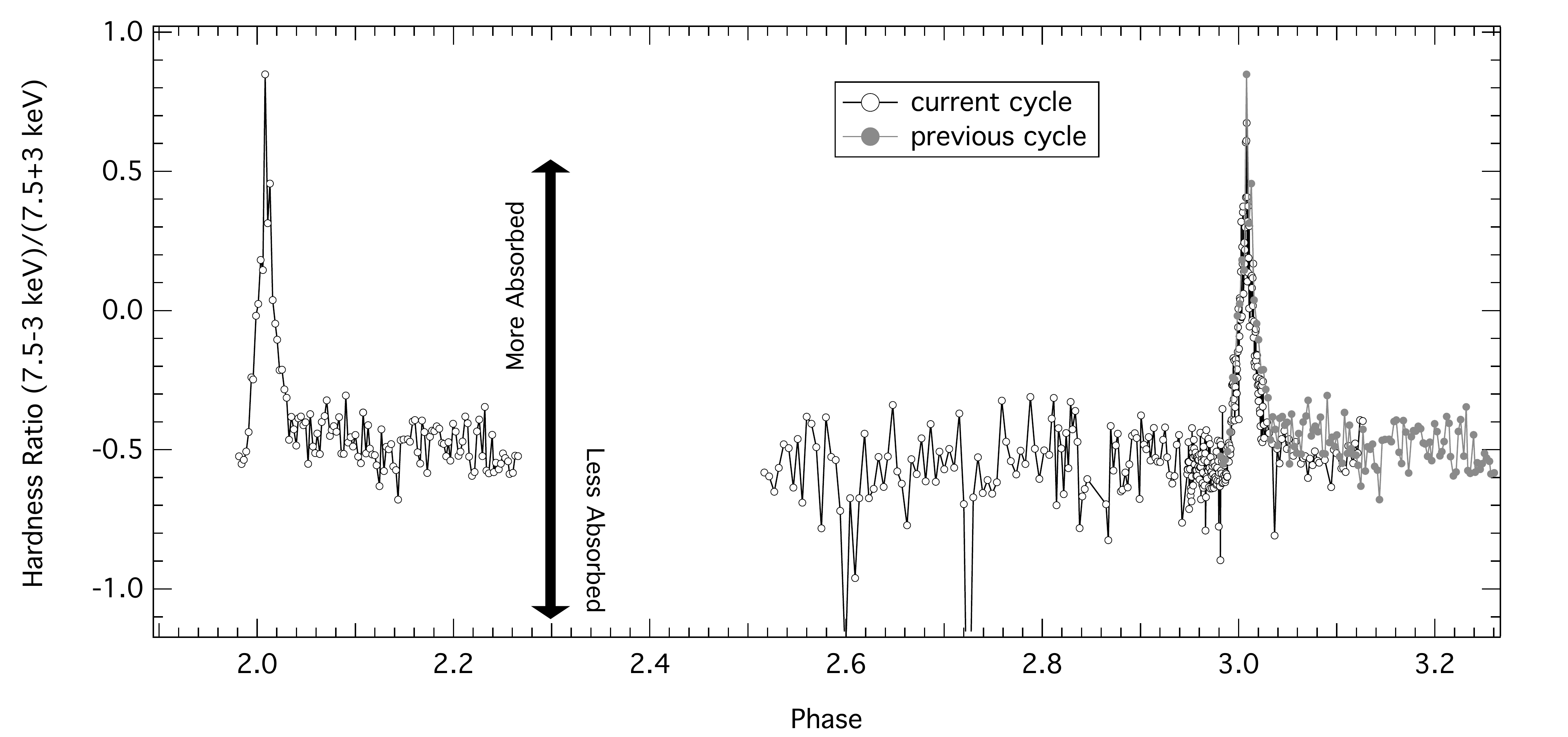} 
   \includegraphics[width=2.5in]{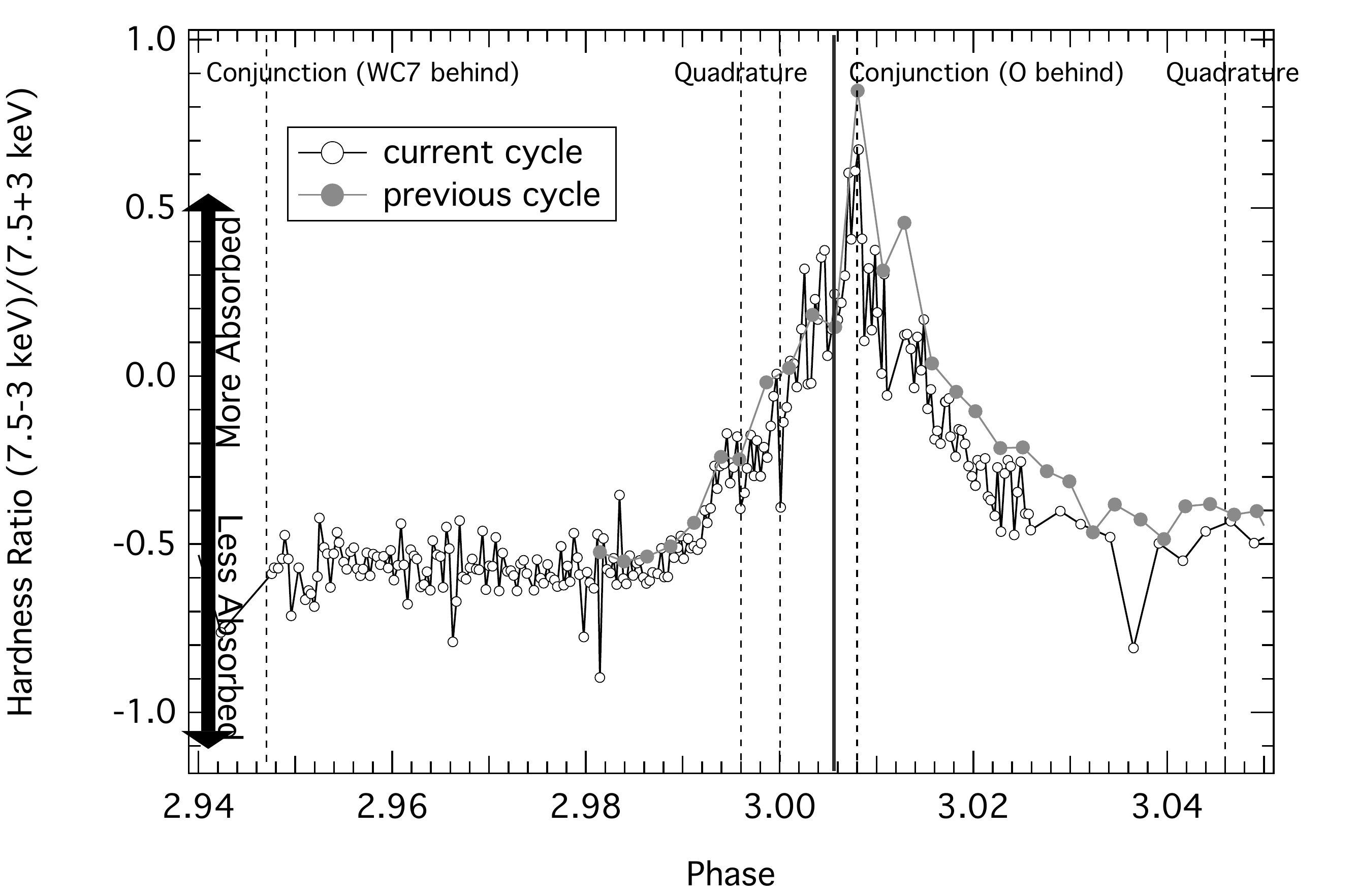} 
   \caption{\footnotesize{WR~140 hardness ratio (HR) variation as measured by the \rxte\ PCA, where HR is the flux at 7.5 keV minus the flux at 3 keV divided by their sum. \textit{Left}: Complete HR curve. \textit{Right}: Detailed view near periastron passage. The thick verticle line just after $\phi=3.0$ marks the phasing of the minimum X-ray brightness. }}
   \label{fig:hr}
\end{figure}

Changes in the X-ray spectrum occur in concert with the observed 2--10 keV flux variations.  These spectral variations have been seen by \exosat, \rosat, and \asca, but the \rxte\ observations provide a more complete view of how the spectrum changes around the orbit.  Because the \rxte\ PCA has rather poor energy resolution, we choose to characterise these spectral variations as changes in X-ray hardness rather than attempting to fully model the spectra. Figure \ref{fig:hr} shows the hardness ratio variations as seen by \rxte, along with a detailed view of the changes near periastron passage.  In this figure the data in grey are data from the previous cycle advanced by 1 period.  In contrast to the X-ray flux variation, the X-ray hardness reaches an extremum at O-star conjunction, as expected if the hardness maximum signals the time of maximum X-ray absorption.    

\section{Modelling}

\begin{figure}[htbp] 
   \centering
   \includegraphics[width=6in]{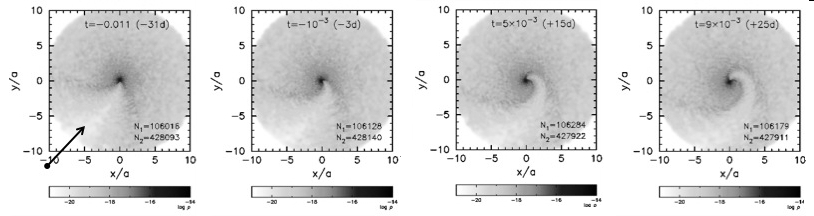} 
   \caption{\footnotesize{A density slice through the orbital plane for a 3-D SPH simulation for WR~140.  Four snapshots are shown, from left to right: near X-ray maximum; near periastron; near X-ray minimum; and near O-star conjunction (maximum X-ray hardness).  The black arrow in the first image shows the observer's line of sight projected onto the orbital plane.}}
   \label{fig:sph}
\end{figure}

The first calculations of the level of X-rays to be expected from the wind-wind shock in colliding wind binaries were by Cherepashchuk (1976) and Prilutskii \& Usov (1976).  Recent modelling efforts have included numerical 3-D hydrodynamic simulations of the wind-wind interaction (Parkin \& Pittard 2008, Okazaki et al. 2008).  Figure \ref{fig:sph} shows a density slice through the orbital plane for a 3-D smooth particle hydrodynamics (SPH) simulation using the Okazaki SPH code (see, for example, Okazaki et al. 2008) for WR~140.  Four snapshots are shown: near X-ray maximum; near periastron; near X-ray minimum; and near O-star conjunction (maximum X-ray hardness).  The axes are marked in units of the orbital semi-major axis $a$. The black arrow in the first image shows the observer's line of sight projected onto the orbital plane.

X-ray maximum occurs when the projected line of sight passes near the middle of the low density region of the O star's wind, channeled by the boundary of the colliding wind shock.  As the stars revolve in their orbit the bow shock around the O star twists behind the companion and gets increasingly obscured by the wind of the WR star through X-ray minimum. At conjunction the X-ray hardness reaches a maximum; this suggests that only the highest-energy photons can penetrate through the thick wind of the WC7 star at this phase. 

\begin{figure}[htbp] 
   \centering
   \includegraphics[width=3in]{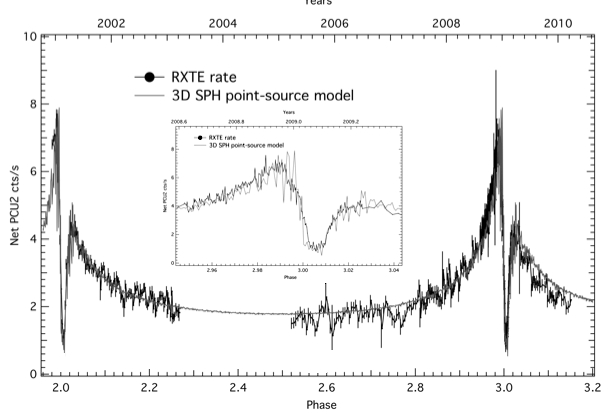} 
   \caption{An X-ray lightcurve from the SPH model compared to the \rxte\ fluxes.}
   \label{fig:sphlc}
\end{figure}

Figure \ref{fig:sphlc} shows an attempt to model the X-ray lightcurve using the SPH model.  The model shown is somewhat artificial in that it assumes a mono-energetic point source of X-rays located at the stagnation point of the flow.  The X-rays produced by this point source are absorbed by the distorted wind structure of the WC7 star which is simulated by the SPH model.  As seen in Fig.~\ref{fig:sphlc} this reproduces quite well the variation seen in the \rxte\ lightcurve.  However, more realistic models under development, in which the X-rays are produced by a distribution of hot gas extended along the wind-wind collision shock, and using a more realistic temperature distribution, do not reproduce the X-ray behaviour as well as the point source model.  This ``distributed-emission'' model is still under investigation. 

\section{Conclusion}

The \rxte\ monitoring of WR~140 confirms that the X-ray emission varies in a predictable way which is mostly consistent over the two orbital cycles studied.  Far from periastron the X-ray emission mostly follows a $1/D$ law expected from an adiabatic shock, though strong deviations are seen near periastron, not all of which can be explained by absorption in the wind of the WC7 star.  Matching the X-ray minimum from 2001 to the 2009 minimum suggests that the X-ray period is $P=2897$ days, about 2 days shorter than the period proposed by Marchenko et al. (2003).  The X-ray data confirm that WR~140 is a good laboratory for the study of the phenomena associated with strong shocks in the astrophysical environment.  

%
\section*{Acknowledgements}
This research was supported through NASA cooperative agreement NNG06EO960A, and made use of the Astrophysics Data System and the HEASARC archive. We express our appreciation to the workshop organisers, and to the  amateur astronomers for their dedication and herculean efforts to observe WR~140 during its 2009 periastron passage.

%
%
\footnotesize
\beginrefer

\refer Bradt, H.~V.,  Rothschild, R.~E.,  Swank, J.~H., 1993, A\&AS, 97, 355

\refer Cherepashchuk, A.~M., 1976, Soviet Astronomy Letters, 2, 138

\refer Marchenko, S.~V., Moffat, A.~F.~J., Ballereau, D., et al., 2003, ApJ, 596, 1295-1304

\refer Okazaki, A.~T., Owocki, S.~P., Russell, C.~M.~P.,  Corcoran, M.~F., 2008, \mnras, 388, L39-L43

\refer Parkin, E.~R., Pittard, J.~M., 2008, \mnras, 388, 1047-1061

\refer Pollock, A.~M.~T., 1985, Space Science Reviews, 40, 63

\refer Prilutskii, O.~F., Usov, V.~V., 1976, Soviet Astronomy, 20, 2

\refer Williams, P.~M., van der Hucht, K.~A. Pollock, A.~M.~T., Florkowski, D.~R., van der Woerd, H.,  Wamsteker, W.~M., 1990, \mnras, 243, 662

\refer White, R.~L., Becker, R.~H., 1995, ApJ, 451, 352

\endrefer           

\end{document}